\documentclass[a4paper]{article}

\usepackage[pages=all, color=black, position={current page.south}, placement=bottom, scale=1, opacity=1, vshift=5mm]{background}
\SetBgContents{
	
}      % copyright

\usepackage[margin=1in]{geometry} % full-width

\usepackage{amsmath}
\usepackage{amsthm}
\usepackage{amssymb}
\usepackage{bbold}
\usepackage{hyperref}
\hypersetup{
	unicode,
	pdfauthor={Shayan Roofeh, Vahid Karimipour},
	pdftitle={The Landau-Streater Channel as a Noisy Quantum Channel},
	pdfsubject={The Landau-Streater Channel as a Noisy Channel Model},
	pdfkeywords={article, template, simple},
	pdfproducer={LaTeX},
	pdfcreator={pdflatex}
}

\usepackage[sort&compress,numbers,square]{natbib}
\theoremstyle{plain}
\newtheorem{theorem}{Theorem}

\theoremstyle{definition}
\newtheorem{definition}{Definition}

\usepackage{graphicx}
\graphicspath{{fig/}}
\usepackage{subcaption}
\usepackage{algorithm, algpseudocode} % use algorithm and algorithmicx for typesetting algorithms
\usepackage{mathrsfs} % for \mathscr command
\usepackage{lipsum}
\usepackage{physics}
\def\be{\begin{equation}}
	\def\ee{\end{equation}}
\def\ba{\begin{eqnarray}}
	\def\ea{\end{eqnarray}}
\def\lo{\longrightarrow}
\def\h{\hskip 1cm }

\def\la{\langle}
\def\ra{\rangle}
\def\a{\alpha}

\def\ni{\noindent}

\def\bex{\begin{dinglist}{110}\dsquare}
	\def\eee{\end{dinglist}}
\def\bet{\begin{dinglist}{110}\bsquare}
	\def\bfr{\begin{mdframed}[backgroundcolor=blue!20]\vspace{0.5cm}}
		\def\efr{\vspace{0.5cm}\end{mdframed}}

\title{The noisy Werner-Holevo channel and its properties}
\author{Shayan Roofeh$^1$ \and Vahid Karimipour$^1$}

\date{
	$^1$\small{Deptartment of Physics, Sharif University of Technology, Tehran, Iran} \\%
}

\begin{document}
	\maketitle

	%\tableofcontents

\begin{abstract}
	
	The interest in the Werner-Holevo  channel $\Lambda_{1}(\rho) =\frac{1}{2}(\tr(\rho)I-\rho^T)$ has been mainly due to its abstract mathematical properties.  We show that in three dimensions and with a slight modification, this channel can be realized as the rotation of qutrit states in random directions by random angles. Our modification takes the form $\Lambda_x(\rho)=(1-x)\rho+x\Lambda_1(\rho)$. Therefore and in view of the potential use of qutrits in quantum processing tasks and their realization in many different platforms, the modified Werner-Holevo channel can be used as a very simple and realistic noise model, in the same way that the depolarizing channel is for qubits. We will make a detailed study of this channel and derive its various properties. In particular, we will use the recently proposed flag extension and other techniques to derive analytical expressions and bounds for the different capacities of this channel. The role of symmetry is revealed in these derivations.  We also rigorously prove that the channel $\Lambda_x$ is anti-degradable and hence has zero quantum capacity, in the region $\frac{4}{7}\leq x\leq 1.$  \\

	\noindent\textbf{Keywords:} Werner-Holevo channel, Classical capacity, Quantum Capacity, Flag-Extension.
\end{abstract}

%\tableofcontents

\section {Introduction}
A quantum state is a matrix that is Hermitian, positive, and of unit trace. Any operation which is quantum mechanically conceivable should preserve these basic properties. Nevertheless the familiar operation of transposing a matrix $\rho\lo \rho^T$,  which for qubits is equivalent to converting any pure state to its orthogonal $|\psi\ra\lo |\psi^T\ra$, while having all these properties is not a feasible quantum operation and cannot be implemented in any physical process. The reason is that it lacks the important and extra property of complete positivity which is required to ensure that local actions on parts of a larger system also retain properties of quantum states.  While transposing a quantum state is forbidden, it was interestingly shown in \cite{WH} that in any dimensions $d$, the so-called Werner-Holevo map
\be
\rho\lo \Lambda_{_{\mathit{WH}}}(\rho)=\frac{1}{d-1}(\tr(\rho)\mathbb{I}-\rho^T),
\ee
despite the presence of the negative sign and the transpose is indeed a completely positive trace-preserving map.  For qubits, it is no surprise that such a map is a physical operation since given a density matrix $\rho=\begin{pmatrix}a&b\\ b^*&c\end{pmatrix}$, we simply have
\be
\Lambda_{_{\mathit{WH}}}(\rho)=\begin{pmatrix}c&-b^*\\ -b&a\end{pmatrix}=\sigma_y\rho \sigma_y,
\ee
which implies that a unitary operation, i.e. a rotation around the $y$ axis by $180^\circ$ covers $\rho$ to $\tr\rho \mathbb{I}-\rho^T.$ For qutrits however, the Werner-Holevo channel is no longer a unitary map, as simply as in the qubit case, but it is still possible to show that it is an acceptable quantum operation. The reason is that it has an explicit Kraus representation which proves that it is indeed a Completely Positive Trace-preserving (CPT) map. In fact, it is known that \cite{pakhomchik_realization_2020}
\be\label{WHLS}
 \Lambda_{_{\mathit{WH}}}(\rho)=\frac{1}{2}(\tr(\rho)\mathbb{I}-\rho^T)=\frac{1}{2}(J_x\rho J_x+J_y\rho J_y+J_z\rho J_z)
\ee
 where $J_x, J_y$, and $J_z$ are the spin-$1$ representations of angular momentum algebra, i.e. 
  \begin{align}\label{KrausLS}
 	J_x &= -i \begin{bmatrix} 0 & 0 & 0 \\ 0 & 0 & 1 \\ 0 & -1 & 0 \end{bmatrix}, &
 	J_y &= -i \begin{bmatrix} 0 & 0 & -1 \\ 0 & 0 & 0 \\ 1 & 0 & 0 \end{bmatrix}, &
 	J_z &= -i \begin{bmatrix} 0 & 1 & 0 \\ -1 & 0 & 0 \\ 0 & 0 & 0 \end{bmatrix}.
 \end{align}
When written in this form, the channel is a special case of the more general types of channels called the Landau-Streater channel \cite{LanS}.
\footnote{
 For higher dimensions,  the Kraus operators are the spin $j$ representation of the angular momentum algebra and the factor $\frac{1}{2}$ should be replaced with $\frac{1}{j(j+1)}$.}
 The importance of the equivalence of the Werner-Holevo with the Landau-Streater channel for qutrits is beyond a simple depiction of the Kraus representation for the former channel. In fact, as has been shown in \cite{LanS}, here we are dealing with the first example of a unital channel which cannot be realized as a collection of random unitary operations. More concretely, the map $\Lambda_{_{\mathit{WH}}}$ while having the property $W_{WH}(I)=I$, cannot be written as $\Lambda_{_{\mathit{WH}}}(\rho)=\sum_i p_i U_i \rho U_i^\dagger$ for any choice of unitary actions and any choice of randomness. This means that the map  $\Lambda_{_{WH}}$ cannot be realized as the random unitary operations (jumps) on the qutrit as it travels along in time or space. It cannot even be written as the convex combination of two other maps. In other words, it is an extreme point in the space of qutrit channels. This is an intriguing result since it is well known that for qubits, any unital map can be written as a random unitary channel \cite{konrad}. Furthermore, the Werner-Holevo channel (\ref{WHLS}) is not continuously connected to the identity channel, i.e. by a parameter which can be tuned to model the effect of environmental noise. \\

\noindent  It is the aim of the present paper to show that by a small modification, that is, by combining it with the identity channel, this map can in fact be a sensible model of noise on qutrits.  
 We will show that the resulting model, defined as, 
\be\label{lambdax}
\Lambda_x(\rho)=(1-x)\rho + x\Lambda_{_{\mathit{WH}}}(\rho)=(1-x)\rho -\frac{x}{2}\rho^T+\frac{x}{2}\tr(\rho)\mathbb{I} \h 0\leq x\leq 1,
\ee
as long as $x\ne 1$, can in fact be represented as a model in which the environmental noise randomly rotates the qutrit by small angles in arbitrary directions, where the parameter $x$ is determined by the probability distribution of random rotations. \\

\noindent This result is significant in several respects: First, from the practical point of view, while universal quantum computing has been conventionally based on qubits, there are many attempts to explore the potential advantages  \cite{caves_qutrit_2000, brus_optimal_2002,molina-terriza_experimental_2005,kendon_bounds_2002, cerf_greenberger-horne-zeilinger_2002,bartlett_quantum_2002, bouda_entanglement_2001} and physical realization of higher dimensional systems, specially qutrits. The latter ranges from orbital angular momentum of light \cite{karimi1, karimi2, karimi3, karimi4} and other photonic platforms \cite{wang_qudits_2020,hugh_trapped-ion_2005,klimov_qutrit_2003,molina-terriza_experimental_2005,bogdanov_qutrit_2004, groblacher_experimental_2006, lanyon_manipulating_2008,schaeff_experimental_2015,babazadeh_high-dimensional_2017,lu_quantum_2020}, to NMR ensembles, \cite{dogra_determining_2014}  superconducting quantum circuits \cite{nature, bianchetti_control_2010,danilin_experimental_2018,yurtalan_implementation_2020,kononenko_characterization_2021}, and trapped ions \cite{klimov_qutrit_2003,randall_efficient_2015,baekkegaard_realization_2019,lindon_complete_2023,low_practical_2020,ringbauer_universal_2022}. 

\noindent Second, the channel $\Lambda_x$ defined in (\ref{lambdax}) plays the same role for qutrits as the depolarizing channel plays for qubits.  It is true that the conventional depolarizing channel for qutrits $\rho\lo (1-p)\rho + \frac{p}{3}I$, allows a random unitary representation in terms of Heisenberg-Weyl operators \cite{wilde_book_2017}. However the Kraus or the error operators of this channel, while being unitary,  are discrete operators which are not connected with the identity operator by a continous parameter representing the level of noise. The level of noise is controlled by the parameter $p$ for all these Kraus operators. Therefore it seems that the channel $\Lambda_x$ shows a more natural type of noise in many practical and numerical studies of  qutrit systems
. In particular, it may have relevance to quantum processes involving the quantum Zeno effect \cite{misra_zenos_1977}, where a series of measurements can slow down the quantum dynamics\cite{raimond_quantum_2012,raimond_phase_2010,bayindir_freezing_2018,bayindir_zeno_2021} or a series of small rotations and measurements can activate bound entanglement \cite{horodecki_bound_1999} in qutrit states \cite{ozaydin_nonlocal_2022}. 
Finally from a mathematical point of view, this result may encourage others to study the vicinity of extreme points of quantum channels in general.   \\

\noindent After proving this physical realization, we will proceed to investigate many of the other properties of the channel. In particular, we will examine the symmetry properties of the channel and the way it affects the most important properties of the channel, namely its various kinds of capacities. As a prerequisite to this investigation, we will study the so-called degradability and anti-degradability \cite{devetak_capacity_2005}conditions for the channel. Equipped with these tools and exploiting the flag extensions \cite{kianvash1, kianvash2} of the channel, we are able to find exact expressions for the Holevo quantity  of the channel, the entanglement assisted capacity of the channel and finally upper and lower bounds for the quantum capacity.  Finally, We  rigorously prove that the channel $\Lambda_x$ is anti-degradable and hence has zero quantum capacity, in the region $\frac{4}{7}\leq x\leq 1.$ 
 \\
 \noindent The structure of this paper is as follows: In section (\ref{randomu}), we show that the channel $\Lambda_x$ is indeed a random unitary channel and show that random rotations of an input qutrit state distort the qutrit in the way suggested by (\ref{lambdax}). 
 In section (\ref{symm}), we study the structural properties of this channel, its spectrum, and its complementary channel. In section (\ref{DegAndAntiDeg}), we extend this study to the important problem of degradability and anti-degradability which is an important requirement for the calculation of various capacities of the channel. Finally, in section (\ref{Capacities}), we use everything we have learned in previous sections to study various capacities of this channel and provide exact expressions or upper and lower bounds for these capacities. Finally in section (\ref{ADD}), we prove that the channel $\Lambda_x$ is anti-degradable and hence has zero quantum capacity, in the region $\frac{4}{7}\leq x\leq 1.$  We conclude the paper with a summary and outlook.   

\section{The random rotation model}\label{randomu}
Consider a qutrit state $\rho$ passing through a communication channel, either in time or in space, and undergoing random kicks (rotations in random directions by random angles). The output state will then be of the form 
\be
\Phi(\rho)=\int d{\bf n}d\theta P({\bf n},\theta)e^{i{\bf n}\cdot {\bf J}\theta}\rho e^{-i{\bf n}\cdot {\bf J}\theta}
\ee
where  $P({\bf n},\theta)$ is the probability density of a rotation being around the direction ${\bf n}$ by an angle $-\frac{\pi}{2}\leq \theta \leq \frac{\pi}{2}$.  We will take this distribution to be uncorrelated in direction and angle. Moreover, in the absence of any preference, we take the distribution in directions to be uniform, hence $P({\bf n},\theta)=\frac{1}{4\pi}P(\theta)$. It is also natural to assume an even distribution function for angles, that is $P(\theta)=P(-\theta)$. We will now use the identity for spin-1 representation (\ref{KrausLS}) 
\be
e^{i{\bf n}\cdot {\bf J}\theta}=1+i\sin\theta \ {\bf n}\cdot {\bf J}+(\cos\theta -1)({\bf n}\cdot {\bf J})^2
\ee
 which in view of $P(\theta)=P(-\theta)$, is simplified to
\be
\Lambda(\rho)=\rho+\la \cos\theta -1\ra\Big[\rho\la{\bf n}\cdot{\bf J} \ra^2+\la {\bf n}\cdot{\bf J}\ra^2 \rho\Big]+\la\sin^2\theta\ra\la {\bf n}\cdot{\bf J} \rho {\bf n}\cdot{\bf J} \ra+\la (\cos\theta-1)^2\ra\la ({\bf n}\cdot{\bf J} )^2\rho ({\bf n}\cdot{\bf J} )^2 \ra
\ee
where $\la f(\theta)\ra:=\int P(\theta)f(\theta)d\theta $ and $\la f({\bf n})\ra:=\frac{1}{4\pi}\int d{\bf n}f({\bf n}) $. We now use the well-known results on the average of unit vectors with uniform distributions to arrive at 
\be
\la ({\bf n}\cdot{\bf J})^2 \ra=\la n_i n_j\ra J_iJ_j=\frac{1}{3}J_iJ_j=\frac{2}{3}\mathbb{I},
\ee
where we have used the identity $J_x^2+J_y^2+J_z^2=2\mathbb{I}$. Moreover, we find
\be
\la {\bf n}\cdot{\bf J} \rho {\bf n}\cdot{\bf J} \ra=\la n_in_j\ra J_i\rho J_j=\frac{1}{3}\sum_i J_i \rho J_i=\frac{2}{3}\Lambda_{_{\mathit{WH}}}(\rho).
\ee
The other term that we should calculate is 
\be\label{nj2}
\la ({\bf n}\cdot{\bf J})^2\rho ({\bf n}\cdot{\bf J} )^2 \ra=\la n_in_jn_kn_l\ra J_iJ_j\rho J_kJ_l.
\ee
Using the identity
\be
\la n_in_jn_kn_l\ra=\frac{1}{15}(\delta_{ij}\delta_{kl}+\delta_{ik}\delta_{jl}+\delta_{il}\delta_{jk})
\ee
and inserting the expression  
$
J_iJ_j=\delta_{ij}\mathbb{I}-|j\ra\la i|
$,  which is obtained from (\ref{KrausLS}), it is straigthforward to simplify the expression (\ref{nj2}). The result is 
\be
\la ({\bf n}\cdot{\bf J})^2\rho ({\bf n}\cdot{\bf J} )^2 \ra=\frac{2}{5}\rho + \frac{1}{15}\rho^T+\frac{1}{15}\tr(\rho)\mathbb{I}.
\ee
Putting everything together, the final result is 
\be\label{Phi}
\Phi(\rho)=a\rho - b\rho^T+\frac{1-a+b}{3}\tr(\rho)\mathbb{I}
\ee
where
\be\label{abc}
a=\frac{1}{15}\la 1+8\cos\theta+6\cos^2\theta\ra,\h b=\frac{1}{15}\la  4+2\cos\theta-6\cos^2\theta \ra
\ee
This shows that random unitary kicks do indeed produce the transpose of $\rho$ with negative sign which is the characteristic of the Werner-Holevo channel, but they also produce the state $\rho$ itself. Note that $b$ is always positive, and $a$ can never be zero, proving that the original Werner-Holevo channel or equivalently the Landau-Streater channel is indeed an extreme point in the space of qutrit channels. In fact the minimum value of $a$ is $\frac{1}{15}=0.067$ which happens when all rotations angles are fixed at $\theta=\pm \frac{\pi}{2}$. This implies that the close vicinity of the channel $\Lambda_{WH}$ is still not representable by random rotations. This of course does not exclude the possibility that this neighborhood be realized by other random unitary ensembles.  The answer to this question is not yet known. When the random angles of rotation are very small so that we can neglect $\la \theta^4\ra$ and higher orders, the channel will take the form
\be
\Phi(\rho)\approx(1-\frac{2}{3}\la \theta^2\ra)\rho-\frac{1}{3}\la \theta^2\ra\rho^T+\frac{1}{3}\la\theta^2\ra \tr(\rho)\mathbb{I},
\ee
which is nothing but the channel $\Lambda_x$  (\ref{WHLS}) with $x= \frac{2}{3}\la \theta^2\ra$.  \\

\noindent We now turn to the study of other properties of our channel and start with a remark on notations and conventions. 

\section{The symmetries of the channel and its spectrum}\label{symm}
We first set up our notations and conventions:\\

\noindent {\bf Notations and conventions:}
Let $H_d$ be a $d-$dimensional Hilbert space; by $L(H_d)$, we mean the space of linear operators, by $L^+(H_d)$, the set of positive semi-definite operators and by $D(H_d)$, the set of unit trace positive operators on $H_d$. $d-$dimensional square matrices are denoted by $M_d$ and $d-$dimensional identity matrix by $I_d$. We employ the notation $\Phi$ to denote an arbitrary quantum channel, which is utilized in various contexts and definitions. For brevity of notations, we sometimes denote a system and its space of operators by the same letter. Therefore, a quantum map $\Phi:A\lo B$ is a shorthand for $\Phi: L(H_A)\lo L(H_B)$. Although the Landau-Streater channel is defined for spin$-j$, in this work, we exclusively study the channel for spin$-1$, which we refer to mostly as the Werner-Holevo channel, since they are equivalent.
The channel $\Lambda_{_{\mathit{WH}}}$ which in the notation used in (\ref{WHLS}) should be denoted by $\Lambda_1$ is obviously covariant in the following sense
\be
\Lambda_{1}(U\rho U^\dagger)=U^*\Lambda_{1}(\rho){U^*}^\dagger,\h \forall\ \ U\in SU(3).
\ee
Moreover, as we will see in section (\ref{DegAndAntiDeg}), it is both degradable and anti-degradable, which implies that it has zero quantum capacity. That is, it cannot convey any quantum information at all. For a detailed treatment, see \cite{filippov_LS}. To see the covariance properties of the channel $\Lambda_x$, we note that since the density matrix $\rho$ transforms as $\rho\lo U\rho U^\dagger$ and $\rho^T$ transforms as $U^*\rho U^T$, the covariance of the channel $\Lambda_x$ reduces from the group $SU(3)$ to its smaller subgroup $SO(3)$,
\be
\Lambda_x(U\rho U^\dagger )=U\Lambda_x(\rho)U^\dagger, \ \ \ \  \ \ \ \ \  \forall U\in SO(3). 
\ee
  
\subsection{Spectral properties of the channel}
The eigenvectors and eigenvalues of the channel can easily be found by noting that for any symmetric matrix, $\rho=\rho^T$ and for any anti-symmetrix matrix $\rho=-\rho^T$. Defining the diagonal matrices, $Z_1=:E_{11}-E_{33}$, $Z_2=:E_{22}-E_{33}$, and the Hermitian matrices,
$X_{sr}:=E_{sr}+E_{sr}$ and $Y_{sr}:=-i(E_{sr}-E_{sr})$ for $s <r$, where $E_{kl}=|k\ra\la l|$ is the matrix with one on its $k-l$ entry and zero otherwise, we find
$$\Lambda_x(I_3)=I_3, \h \Lambda_x(Z_1)=(1-\frac{3x}{2})Z_1, \h \Lambda_x(Z_2)=(1-\frac{3x}{2})Z_2,$$
\be
\h \Lambda_x(X_{sr})=(1-\frac{3x}{2})X_{sr},\h \Lambda_x(Y_{sr})=(1-\frac{x}{2})Y_{sr}.
\ee
Counting the degeneracies of the eigenvalues, this shows that the determinant of the channel is equal to  
\be
Det(\Lambda_x)=(1-\frac{x}{2})^3(1-\frac{3}{2}x)^5.
\ee
This shows that the determinant of the channel is negative in a certain range, namely
\be
Det(\Lambda_x)<0,\h \frac{2}{3}<x\leq 1.
\ee
In this range, the channel is not infinitely divisible and does not represent a Markovian evolution \cite{wolf_dividing_2008}. The question remains whether, within the range of $0\leq x\leq \frac{2}{3}$, the channel can be described as the exponential of Lindbladian dynamics. We will explore this question in our future work. 

\subsection{Complementary Channel}
\label{Comp-channel}
\subsubsection{Definition and Covariance of Complementary Channel}

\iffalse calculation of many types of capacities reduces to optimization of quantities related to both the channel and its complement. Therefore, in this subsection, we review the concept of the complement of a channel and investigate how the covariance and symmetry properties of a channel are reflected in its complement. \fi
The concept of the complement of a channel hinges on the well-known Stinespring's dilation theorem  \cite{stinespring}, which states that a quantum channel $\Phi: A\lo B$ can be constructed as a  unitary map $U:A\otimes E\lo B\otimes E'$, where $E$ and $E'$ are the environments of $A$ and $B$ respectively. More formally, we have 
\begin{equation}
\Phi(\rho) = \text{tr}_{E'}(U\rho U^{\dagger}),
\end{equation}
where $U$ denotes an isometry mapping from $A$ to ${B} \otimes E'$. In this configuration, the complementary channel $\Phi^c: A \longrightarrow E'$ is defined by:
\begin{equation} \label{steincomp}
\Phi^c(\rho) = \text{tr}_{B}(U\rho U^{\dagger}),
\end{equation}
Constituting a mapping from the input system to the output environment (see figure (\ref{fig:comp})). It's important to note that the complement of a quantum channel is not unique, but there exists a connection between them through isometries, as detailed in \cite{datta_complementarity_2006}. The Kraus operators of the channel $\Phi$ and its complement $\Phi^c$ are related as follows
\cite{smaczynski2016selfcomplementary}: 
\begin{equation} \label{krauscomp}
	\begin{split}
		&\Phi(\rho)= \sum_\alpha K_{\alpha} \rho K_{\alpha} ^{\dagger}, \\
		&\Phi^c(\rho)= \sum_i R_i \rho R_i ^{\dagger}, \\
		&(R_i)_{\alpha,j}= (K_{\alpha})_{i,j}.
	\end{split}     
\end{equation}

\noindent \\As shown in \cite{poshtvan_capacities_2022}, the covariance of a channel also induces a covariance on the complementary channel. The theorem of \cite{poshtvan_capacities_2022} states that if a channel $\Phi$ is covariant in the form
\be
\Phi\big[U(g)\rho U^\dagger(g)\big]=V(g)\Phi(\rho)V^\dagger(g)\h \forall \ g\in G,
\ee
where $U(g)$ and $V(g)$ are two representations of the group element $g$ in a group $G$, then the complement channel is covariant in the following form
\be
\Phi^c\big[U(g)\rho U^\dagger(g)\big]=\Omega^\dagger (g)\Phi^c(\rho)\Omega(g)\h \forall \ g\in G,
\ee
where $\Omega(g)$ is the representation defined in the following form
\be\label{cov1}
V^\dagger(g)K_i U(g)=\sum_{j}\Omega_{i,j}(g)K_j.
\ee

\subsubsection{Complementary Channel of $\Lambda_x$}
The formula (\ref{krauscomp}) gives a very simple recipe for writing the Kraus operators of the complementary channel easily. Put the first rows of all the Kraus operators in consecutive rows of a matrix and call it $R_1$, put the second rows of all the Kraus operators in consecutive rows of a matrix and call it $R_2$, and so on and so forth. 
Since the Kraus operators of the channel $\Lambda_x$ are given by 

\be
K_0=\sqrt{1-x}\begin{bmatrix} 1 & 0 & 0 \\ 0 & 1 & 0 \\ 0 & 0 & 1 \end{bmatrix}, 
\ee
and
\begin{align}
	\label{KrausLSS}
	K_1 &= -i  \sqrt{\frac{x}{2}}\begin{bmatrix} 0 & 0 & 0 \\ 0 & 0 & 1 \\ 0 & -1 & 0 \end{bmatrix}, &
	K_2 &= -i \sqrt{\frac{x}{2}}\begin{bmatrix} 0 & 0 & -1 \\ 0 & 0 & 0 \\ 1 & 0 & 0 \end{bmatrix}, &
	K_3 &= -i  \sqrt{\frac{x}{2}}\begin{bmatrix} 0 & 1 & 0 \\ -1 & 0 & 0 \\ 0 & 0 & 0 \end{bmatrix},
\end{align}
it is readily found that the Kraus operators of $\Lambda^c_x$ are given by 
\begin{equation}\label{Rs}
	R_1= \begin{bmatrix}
		\sqrt{1-x} & 0 & 0 \\
		0 & 0 & 0 \\
		0 & 0 & i\sqrt{x/2} \\
		0& -i\sqrt{x/2} & 0\\
	\end{bmatrix}
	,\:\:\:\: R_2= \begin{bmatrix}
		0 & \sqrt{1-x} & 0 \\
		0 & 0 & -i\sqrt{x/2} \\
		0 & 0 & 0\\
		i	\sqrt{x/2} & 0 & 0 \\
	\end{bmatrix}
	,\:\:\:\:
	R_3= \begin{bmatrix}
		0 & 0 & \sqrt{1-x} \\
		0 & i\sqrt{x/2} & 0 \\
		-i\sqrt{x/2} & 0 &0\\
		0 & 0 & 0\\
	\end{bmatrix}.
\end{equation}
In passing and for future use, it is instructive to note the effect of the complementary channel on a general matrix
\be
X=\left(
\begin{array}{ccc}
	a & b & c \\
	d & e & f \\
	g & h & k \\
\end{array}
\right),
\ee
which is readily found from (\ref{Rs}) to be
\be\Lambda_x^c(X)=\frac{1}{2}
\left(
\begin{array}{cccc}
	2(1-x) ((a+e+k)) & {i \sqrt{2x(1-x) } (f-h)} & -{i \sqrt{2x(1-x) } (c-g)} & {i \sqrt{2x(1-x) } (b-d)} \\
	{i \sqrt{2x(1-x) } (f-h)} & x (e+k) & - d x & - g x \\
	-{i \sqrt{2x(1-x) } (c-g)} & - b x & x (a+k) & - h x \\
	{i \sqrt{2x(1-x) } (b-d)} & - c x & - f x &  x (a+e) \\
\end{array}
\right).
\label{Comp:explicit}
\ee

\iffalse
\be
\Lambda(I)=3(1-x)|0\ra\la 0|\oplus x I,\ \ \  \ \ \Lambda(X_{ij})=-\frac{x}{2}X_{ij},\ \ \ \ \  \Lambda(Y_{ij})=i\sqrt{\frac{x(1-x)}{2}}(|0\ra\la \epsilon_{ij}|+|\epsilon_{ij}\ra\la 0|)\oplus \frac{x}{2}Y_{ij}
\ee
where
\be
|\epsilon_{12}\ra=|3\ra,\ \ \ |\epsilon_{23}\ra=|1\ra, \ \ \ |\epsilon{31}\ra=|2\ra.
\ee
\fi
\noindent In the case of the channel $\Lambda_x$, we have
\be
U^\dagger K_0U=K_0, \quad U^\dagger K_iU=\sum_j U_{ij} K_j, \quad U\in SO(3),
\ee
where $U=e^{i\theta{\bf n}\cdot {\bf J}}$ is the spin-$1$ representation of a rotation. Hence
\be
\Omega=\begin{pmatrix}1&{\bf 0}^T\\ {\bf 0}& U\end{pmatrix}.
\ee
Therefore, the channel $\Lambda^c_x$ is also covariant under $SU(3)$ when $x=0$ and is covariant under $SO(3)$ for arbitrary $x$.

\section{Degradability and Anti-Degradability}\label{DegAndAntiDeg}
Degradable and anti-degradable channels belong to a noteworthy class of completely positive trace-preserving maps. These channels possess advantageous properties that we use in this paper to analyze various capacities. 

\begin{figure}[H]
	\centering
	\includegraphics[width=11cm]{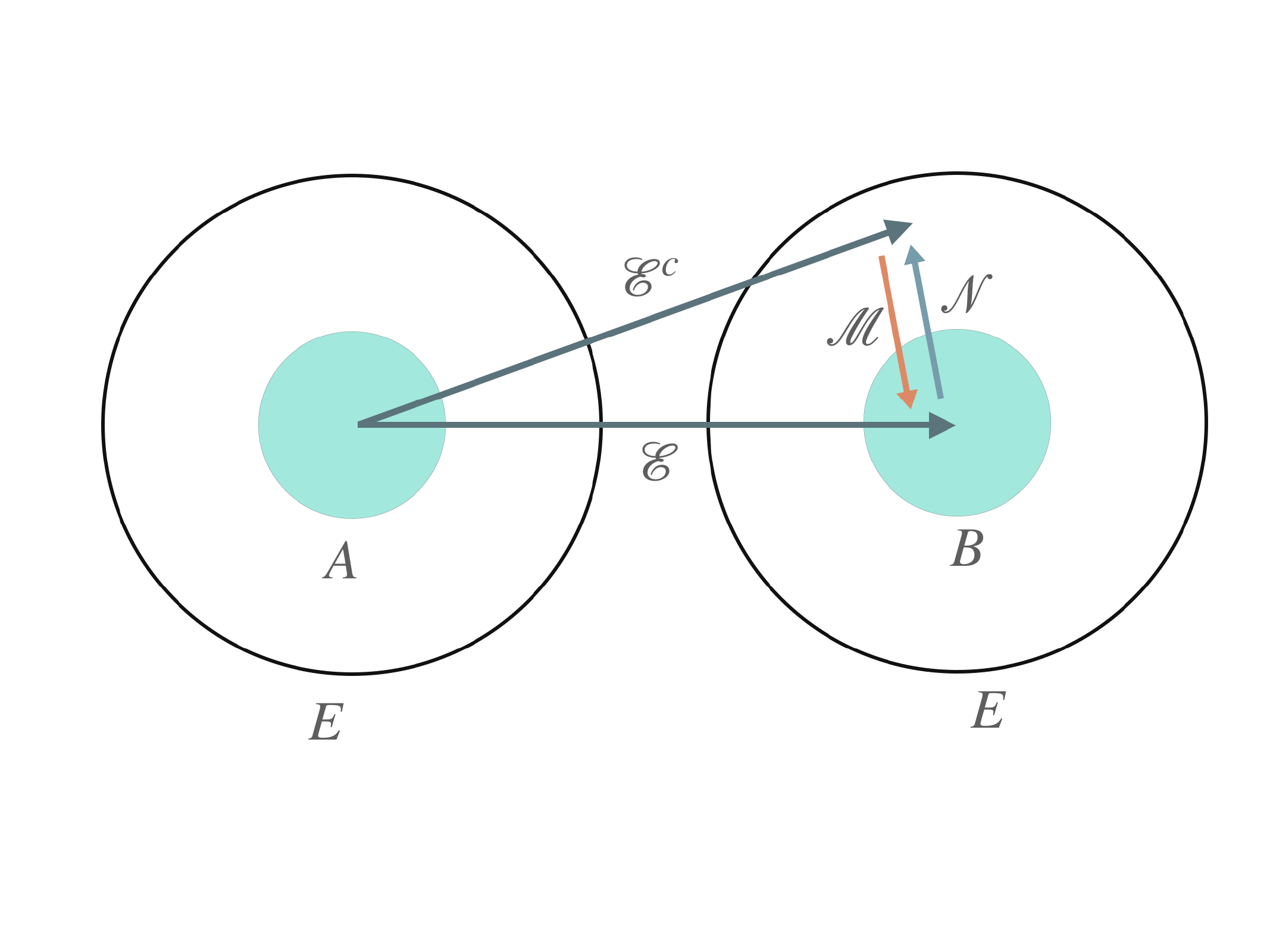}\vspace{-1cm}
	\caption{The channel $\mathcal{E}$, its complement $\mathcal{E}^c$ and the channels $\mathcal{M}$ and $\mathcal{N}$, in the definition (\ref{definitonofdeg}) of degradibility and anti-degradibility. For simplicity, we have taken the environments of $A$ and $B$, the same as $E$. In principle, they can be different.}
	\label{fig:comp}
\end{figure}

\begin{definition} \cite{devetak_capacity_2005,cubitt_structure_2008}
\label{definitonofdeg}
 Let $\mathcal{E}:A\lo B$ and $\mathcal{E}^c:A\lo E$ be a quantum channel and its complement respectively. Then  $\mathcal{E}$ is degradable if there exist a quantum channel $\mathcal{N}:B\lo E$, such that  $\mathcal{N}\circ \mathcal{E}(\rho)=\mathcal{E}^c(\rho)$. It is anti-degradable if  there exist a quantum channel $\mathcal{M}:E\lo B$ such that $\mathcal{M}\circ \mathcal{E}^c(\rho))=\mathcal{E}(\rho)$, figure (\ref{fig:comp}).
\end{definition}

\noindent We first note that in the specific point $x=1$, where we have the pure Landau-Streater channel and $K_0=0$, $K_{1,2,3}=J_{x,y,z}$, it is readily seen that the Kraus operators $R_i$ will be $3$ by $3$ matrices and  

\be
R_1=-K_1, \ \ \ R_2=-K_2,\ \ \ R_3=-K_3.
\ee
That is, the Landau-Streater channel and its complement are the same, 
\be
\label{LS-DAD}
\Lambda_{1}(\rho)=\Lambda^c_{1}(\rho).
\ee
This implies that the Landau-Streater channel is both degradable and anti-degradable simultaneously and has important implications for the quantum capacity of the channel, as we will later explain. 
In the other extreme, when $x=0$ and we have the identity channel, 
$
\Lambda_0(\rho)=\rho,
$
we find
\be
R_1=\begin{pmatrix}1,0,0\end{pmatrix}=|0\ra\la 1|,\h R_2=\begin{pmatrix}0,1,0\end{pmatrix}=|0\ra\la 2|,\h R_3=\begin{pmatrix}0,0,1\end{pmatrix}=|0\ra\la 3|,
\ee
which leads to 
\be
\Lambda_0^c(\rho)=\tr(\rho)|0\ra\la 0|.
\ee
It is then obvious that the channel is degradable with 
$\mathcal{N}(\rho)=\tr(\rho)|0\ra\la 0|$, since
$\Lambda_0^c(\rho)=(\mathcal{N}\circ \Lambda)(\rho)$ and since no channel can retrieve a state from its trace, the channel 
$\Lambda_x$
is not anti-degradable at point $x=0$.

\section{Capacities of the noisy Landau-Streater Channel}\label{Capacities}
We have now concluded our partial investigation of the properties of the channel $\Lambda_x$. In this new section, our focus shifts to the examination of its various capacities. Calculating different capacities analytically is not always feasible, except in some special cases \cite{king_capacity_2003,bennett_capacities_1997,smith_quantum_2008,ouyang_channel_2014,kianvash_bounding_2022,chessa_quantum_2021,arqand_quantum_2020,oskouei_capacities_2022,leditzky_quantum_2018,chessa_resonant_2023,chessa_partially_2021}. We leverage the symmetry properties of this channel to derive analytical expressions for its Holevo quantity and entanglement-assisted capacity. We then numerically find upper and lower bounds for the quantum capacity and determine the region of non-zero quantum capacity. This is shown in figure (\ref{fig:Q-capacity}). Let us start with the classical capacity. 
\subsection{Classical Capacity}
The classical capacity represents the ultimate rate at which classical messages can be faithfully transmitted through a channel upon being encoded into quantum states. This concept is formulated as \cite{schumacher1997sending,holevosignal}:
\begin{equation} \label{classicalcap}
	\begin{split}
		C_{cl}(\Phi)= \lim_{n\longrightarrow \infty} \frac{1}{n} \chi^*(\Phi^{\otimes n}),
	\end{split}    
\end{equation}

\noindent where $\chi^*(\Phi)$ is defined as:

$$\chi^*(\Phi)= \max_{p_i, \rho_i} \ S\left(\sum_i p_i\Phi(\rho_i)\right)-\sum_i p_iS\left(\Phi(\rho_i)\right).$$

\noindent In this context, the von-Neumann entropy is represented as $S(\rho)=-\text{Tr}(\rho\log \rho)$ \cite{nielsen}. It's crucial to note that $\chi^*$ exhibits superadditivity, meaning that $n\chi^*(\Phi)\leq \chi^*(\Phi^{\otimes n})$. This necessitates the regularization step in equation (\ref{classicalcap}) for calculation of the classical capacity  \cite{hastings_superadditivity_2009}. 
\subsubsection{The Holevo quantity}
This regularization being extremely difficult, we, as many others \cite{filippov_LS,darrigo_classical_2013, giovannetti_information-capacity_2005}, try to calculate the Holevo quantity which is a lower bound on the regularized classical capacity, that is

\be
 \chi^*(\Phi)\leq C_{cl}(\Phi).
\ee
Since the channel $\Lambda_x$ is irreducibly covariant, we can use a result of Holevo \cite{holevo_remarks}, according to which $\chi^*(\Lambda_x)$ is given by 

\begin{equation}
	\label{irreducibly-covariant-Holevo}
	\chi^*(\Lambda_x)= \log d - \min_{\rho} S\left( \Lambda_x(\rho)\right).
\end{equation}
Thus, our task reduces to finding the minimum output entropy state of the channel $\Lambda_x$. As entropy is a concave function, we know that this minimum output state can be taken to be a pure state \cite{wilde_book_2017}. To find this minimum output entropy state, we use the covariance properties of the channel. First, consider the two endpoints of the parameter space $\{0,1\}$, namely $x=0$ and $x=1$. In these two points, the channel has $SU(3)$ covariance, which means that if a given state $|\psi_0\ra$ is a minimum output entropy state, so is the state $U|\psi_0\ra$ where $U$ is an arbitrary $U\in SU(3)$. This is due to the fact that
\be
S(\Lambda_i(U\rho U^\dagger) )=S(U\Lambda_i(\rho) U^\dagger)=S(\Lambda_i(\rho)), \h i=0,1.
\ee
We can now use this full $SU(3)$ covariance to choose the minimum output entropy state as 
\be
\rho_{min}=|1\ra\la 1|,\h {\rm where}\h  |1\ra=\begin{pmatrix}1\\ 0 \\ 0 \end{pmatrix}.
\ee
This immediately leads to 
\be
\Lambda_{0}(|1\ra\la 1|)=\begin{pmatrix}1&0&0\\ 0&0&0\\0&0&0\end{pmatrix},\h \Lambda_{1}(|1\ra\la 1|)=\begin{pmatrix}0&0&0\\ 0&1&0\\0&0&1\end{pmatrix},
\ee
from which we obtain
\be
\chi^*(\Lambda_{0})=\log_23,\h \chi^*(\Lambda_{1})=\log_23-1.
\ee
However, when $x$ is not an endpoint, the channel's full $SU(3)$ covariance is broken down to its $SO(3)$ subgroup. It is no longer possible to transform any arbitrary state in the complex Hilbert space $H_3$ to a fixed state like $|1\ra$ by $SO(3)$ rotations. To use this limited covariance, suppose that the minimum output entropy  state is given by
\be
|\Psi\ra=\begin{pmatrix}\psi_1\\ \psi_2\\ \psi_3\end{pmatrix},
\ee
where $\psi_1,\psi_2$ and $\psi_3$ are complex numbers, modulo the normalization condition and a global phase. We now use a group element $O_1\in SO(3)$ with a suitable rotation parameter $\a$ 
\be
O_1=\begin{pmatrix}\cos\alpha&\sin\alpha&0 \\ -\sin\alpha&\cos\alpha&0\\0&0&1\end{pmatrix},
\ee
to make $\psi_2$ real. (i.e. we can always choose $\a$ so that $Im(-\psi_1\sin\a +\psi_2\cos\a )=0$) and turn $|\Psi\ra$ to
\be
|\Psi\ra=\begin{pmatrix}\psi_1\\ r_2\\ \psi_3\end{pmatrix},\h r_2\in R. 
\ee
With another rotation of the form 
\be
O_2=\begin{pmatrix}\cos\beta&0&\sin\beta \\ 0&1&0\\-\sin\beta&0&\cos\beta\end{pmatrix},
\ee
with a suitable parameter $\beta$, we can make $\psi_3$ also real 
\be
|\Psi\ra=\begin{pmatrix}\psi_1\\ r_2\\ r_3\end{pmatrix}, \h r_2,\ r_3\in R.
\ee
Finally we use the rotation matrix $O_3\in SO(3)$ 
\be
O_3=\begin{pmatrix}1&0&0\\ 0&\cos\gamma&\sin\gamma \\ 0&-\sin\gamma&\cos\gamma\end{pmatrix}.
\ee
To eliminate the third component ($-r_2\sin \gamma +r_3\cos\gamma =0$) and set the form of $|\Psi\ra$ into
\be
|\Psi\ra=\begin{pmatrix}\cos\theta  e^{i\phi}\\ \sin\theta\\ 0\end{pmatrix},
\ee
where normalization has also been used. The output of the channel $\Lambda_x$ is now given by
\be
\Lambda_x(|\Psi\ra\la \Psi|)=\begin{pmatrix}(1-x)\cos^2\theta + \frac{x}{2}\sin^2\theta & \cos\theta\sin\theta \big[(1-x)e^{i\phi}-\frac{x}{2} e^{-i\phi}\big]\\  \cos\theta\sin\theta \big[(1-x)e^{i\phi}-\frac{x}{2} e^{-i\phi}\big]& (1-x)\sin^2\theta + \frac{x}{2}\cos^2\theta \end{pmatrix}\oplus \frac{x}{2}I_1.
\ee
Let us abbreviate the above matrix to 
$\Lambda_x(|\Psi\ra\la \Psi|)=M\oplus \frac{x}{2}I_1$. Then, the eigenvalues of this matrix are $\frac{x}{2}$ plus the two eigenvalues of the matrix $M$. Instead of explicit calculation of the eigenvalues of $M$ which we denote as $\lambda_1$ and $\lambda_2$, we note that 

\be
\tr(M)\equiv \lambda_1+\lambda_2=1-\frac{x}{2},\h Det(M)\equiv \lambda_1\lambda_2=\frac{1}{2}x(1-x)(1-\sin^2 2\theta \sin^2\phi).
\ee
The sum of eigenvalues is fixed and independent of the input state $|\Psi\ra. $ The minimum output entropy is obtained when the two eigenvalues are as far apart as possible, i.e., when $Det(M)$ is a minimum that is realized for  
\be
\theta=\frac{\pi}{4},\h \phi=\frac{\pi}{2}.
\ee
This leads to $Det(M)=0$. Thus, the minimum output entropy state is 
\be
|\Psi\ra=\frac{1}{\sqrt{2}}\begin{pmatrix}i\\1\\0\end{pmatrix},
\ee
and the eigenvalues of $M$ are $\{\lambda_1=0,\ \lambda_2=1-\frac{x}{2}\}$. Taking into account the third eigenvalue $\lambda_3=\frac{x}{2}$
the minimum output entropy will be 
\be
S=-\frac{x}{2}\log\frac{x}{2}-(1-\frac{x}{2})\log (1-\frac{x}{2}),
\ee
which leads to 
\be
\chi^*(\Lambda_x)=\log_23 +\frac{x}{2}\log\frac{x}{2}+(1-\frac{x}{2})\log (1-\frac{x}{2}),
\label{Classical-Capacity}
\ee
which, in view of the (\ref{Classical-Capacity}), shows that the classical capacity is a continuous function of the parameter $x$. Figure (\ref{fig:CIEN}) shows the Holevo quantity as  a function of $x$. 
\subsubsection{Upper bound for classical capacity}
We employ an upper bound using semi-definite programming, as introduced in reference \cite{wang_classical} and denoted as $C_\beta$. Consider a quantum channel $\Phi: A \rightarrow B$, where $J(\Phi) = \sum_{ij} (\dyad{i}{j}) \otimes (\Phi(\dyad{i}{j}))$ to represent its Choi matrix \cite{choi1975}. Take $S_B$ and $R$ as hermitian matrices in $B$ and $A\otimes B$ respectively. Then it is shown that \cite{wang_classical}:

\[ C_{cl}(\Phi) \leq C_\beta(\Phi) := \log\left(\min_{\Delta} \:\Tr(S_B) \right) \]

\noindent where the subscript $\Delta$ is meant to denote the constraints on $S_B$ in this optimization which is to be performed over all feasible hermitian matrices $S_B$ and $R$, subject to:

\[ -R \le J(\Phi)^{T_B} \le R \]
\[ -I_A\otimes S_B \le R^{T_B} \le I_A \otimes S_B \]

\begin{figure}[H]
  \centering
    \includegraphics[width=\textwidth]{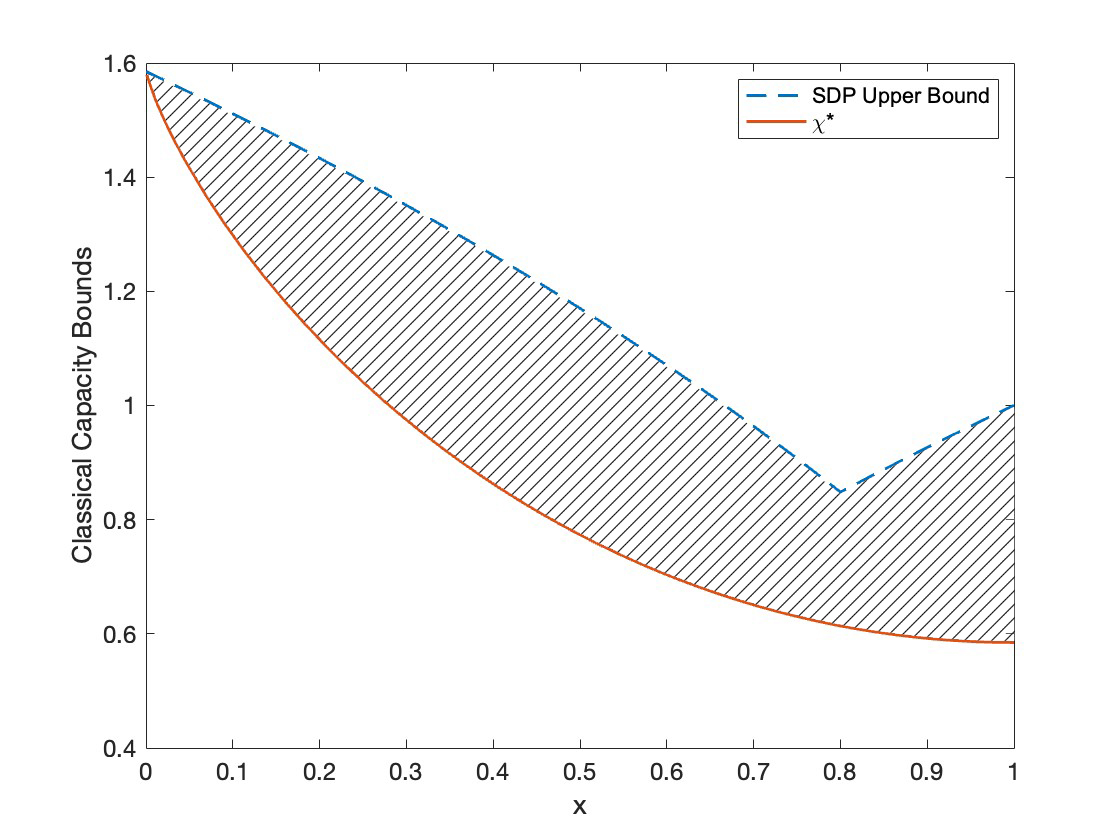}
    \caption{Color-Online: The Holevo quantity $\chi^*$ and classical capacities' SDP upper bound of the channel $\Lambda_x$ as a function of $x$.  The shaded region shows the allowable region for the classical capacity of the channel. When $x=0$, i.e. the identity channel, the upper and lower bounds are equal to $\log_2(3)$. The parameter $x$ and the quantity $\chi^*$ are dimensionless. The parameter $x$ and the capacities are dimensionless.}
    \label{fig:C-capacity}
  \end{figure}

\noindent Here, $T_B$ signifies the partial transpose operation with respect to subspace $B$, defined as $(\dyad{ij}{kl})^{T_B} = \dyad{il}{kj}$. We utilize this method to establish an upper bound for the classical capacity of the channel (\ref{lambdax}), as visually represented in figure \ref{fig:C-capacity}.

\subsection{Entanglement-Assisted Capacity}
Entanglement-assisted capacity is a measure of the maximum rate at which quantum information can be transmitted through a noisy quantum channel when the sender and receiver are allowed to share an entangled quantum state \cite{bennett_entanglement-assisted_1999}. The entanglement-assisted classical capacity of a channel $\Phi$ is determined by \cite{bennett_entanglement-assisted_2002}:
\begin{equation}
	C_{ea}(\Phi)=\max_{\rho} I(\rho,\Phi),
\end{equation}
where 
\be
I(\rho, \Phi):=S(\rho)+S(\Phi(\rho))-S(\rho,\Phi).
\ee

\noindent Here $S(\rho,\Phi)$ is the output entropy of the environment, referred to as the entropy exchange \cite{nielsen_1998}, and is represented by the expression $S(\rho,\Phi) = S({\Phi}^c_x(\rho))$ where $\Phi^c_x$ is the complementary channel (see section (\ref{Comp-channel})).
According to Proposition 9.3 in \cite{holevoBook}, the maximum entanglement-assisted capacity of a covariant channel $\Phi$ is attained for an invariant state $\rho$. In the special case where $\Phi$ is irreducibly covariant, the maximum is attained on the maximally mixed state. Hence, for the channel (\ref{lambdax}):

\begin{equation}
	C_{ea}(\Lambda_x)=S(\frac{I}{3})+S(\Lambda_x(\frac{I}{3}))-S({\Lambda_x}^c(\frac{I}{3})),
\end{equation}
which, given the unitality of the channel, leads to
\begin{equation}
	C_{ea}(\Lambda_x)=2S(\frac{I}{3})-S({\Lambda_x}^c(\frac{I}{3}))=2\log_23-S({\Lambda_x}^c(\frac{I}{3}).
\end{equation}
From (\ref{Comp:explicit}), one finds
$$
{\Lambda}_x^c (\frac{I}{3}) = \begin{pmatrix}
	1-x & 0 & 0 & 0  \\
	0 & \frac{x}{3} & 0 & 0  \\
	0 & 0 & \frac{x}{3} & 0  \\
	0 & 0 & 0 & \frac{x}{3}  \\
	\end{pmatrix},
$$
which gives the final expression for the entanglement-assisted classical capacity 
\be
C_{ea}(\Lambda_x)=2\log_23+x\log_2\frac{x}{3}+(1-x)\log_2(1-x).
\label{EACapacity}
\ee
\begin{figure}[H]
	\centering
	\includegraphics[width=\textwidth]{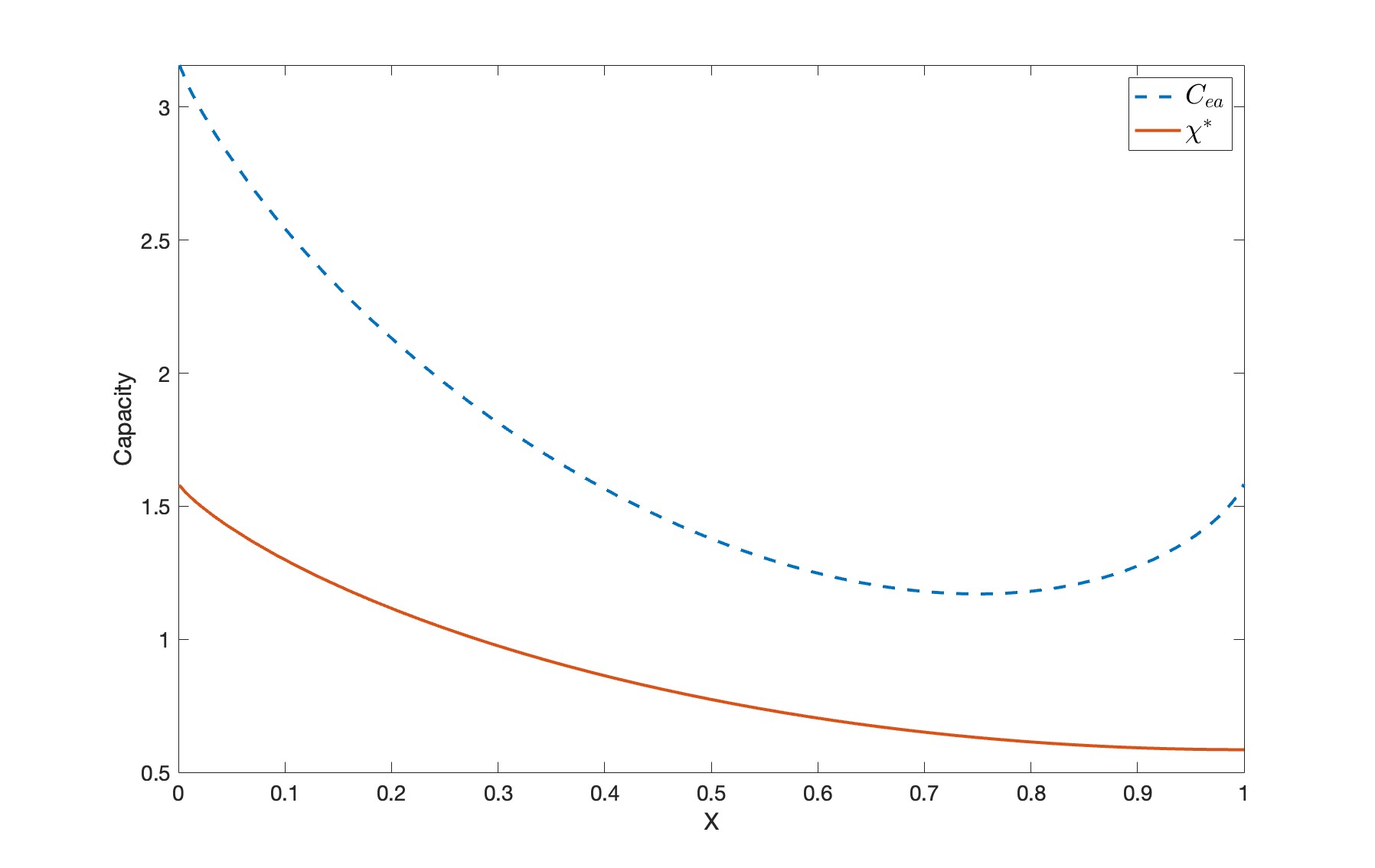}
	\caption{ Color-Online: The Holevo quantity $\chi^*$ and Entanglement Assisted ($C_{ea}$)  capacites of the channel $\Lambda_x$ as a function of $x$. All quantities and parameters are dimensionless.
	}
	\label{fig:CIEN}
\end{figure} 
\noindent The specific limiting points of the graph (\ref{fig:CIEN}) can be understood on physical grounds as follows. First, consider the point $x=0$, where the channel is identity. In this case, without using entanglement, we can communicate the classical bits $0,\ 1$ and $2$, by encoding them into the qutrit states $|0\ra, |1\ra$ and $|-1\ra$, which after passing through the channel can be retrieved in a safe form. This will give mutual information of 
\be
I_{cl, x=0}(X;Y)\equiv H(X)-H(X|Y)=\log 3-0=\log 3\approx 1.58.
\ee
When $x=0$, we can run the usual dense-coding protocol and communicate two classical trits by sending each single qutrit through the channel. This gives the mutual information $I_{ea, x=0}=H(X)-H(X|Y)=\log 9=2\log 3\approx 3.17 $ as shown in figure (\ref{fig:CIEN}). At the other endpoint, when $x=1$, we calculate the mutual information as follows. Let us  use the same encoding $(0,\ 1,\ 2)\lo |0\ra,\ |1\ra, \ |-1\ra)$. Call an arbitrary qutrit state $|m\ra$ and send it through the channel. This turns into the mixed state $$\Lambda_{1}(|m\ra\la m|)=\frac{1}{2}(I-|m\ra\la m|),$$ and is received by the receiver. A projective measurement by the receiver leads to the following pattern of conditional probabilities, 
$P(y=m|x=m)=0$ and $P(y\ne m|x=m)=\frac{1}{2}$. This gives the following mutual information
\be
I_{cl, x=1}(X:Y)=H(X)-H(X|Y)=\log 3- 1\approx 0.58.
\ee
It remains to make a concrete calculation of $C_{ea}$ at $x=1$. This is the most non-trivial, and we have only found examples that are compatible with the results in figure (\ref{fig:CIEN}). In appendix (\ref{ProtocolsforEAC}), we explain two different protocols that yield different mutual information, both compatible with the result (\ref{EACapacity}).
Figure (\ref{fig:CIEN}) shows both classical and entanglement-assisted capacity as a function of $x$.

\subsection{Quantum Capacity}
The largest rate at which quantum information can be sent reliably through a quantum channel $\Phi$ is represented by its quantum capacity $Q(\Phi)$. The coherent information $I_c(\Phi,\rho)=S(\Phi(\rho))-S({\Phi}^c(\rho))$ is an entropic quantity that can be used to express the quantum capacity. The maximum value of $I_c$ over all input states $\rho$ is denoted by $Q_1(\Phi)$, i.e:

\begin{equation}
Q_1(\Phi)= \max_\rho I_c(\Phi,\rho)= \max_\rho S(\Phi(\rho))-S({\Phi}^c(\rho)).
\end{equation}
 To define the quantum capacity of a channel, $Q(\Phi)$, we use the single-letter quantum capacity and take the limit as $n$ approaches infinity\cite{barnum_information_1998, lloyd_capacity_1997,shor_quantum_2002,devetak_private_2005}:

\begin{equation}
\label{regularizing_quantum_capacity}
 \lim_{n\rightarrow \infty} \frac{1}{n}Q_1(\Phi^{\otimes n})
 \end{equation}.

\noindent When the channel $\Phi$ is degradable, this regularization is not necessary, and we have  \be Q(\Phi)=Q_1(\Phi),\ee calculation of $Q_1(\Phi)$ is then a convex optimization problem. In general, however,  performing the regularization in (\ref{regularizing_quantum_capacity}) is extremely difficult, if not impossible \cite{cubitt_2015}; even for very simple qubit channels like the depolarizing or the Pauli channel, determining the quantum capacity remains elusive \cite{wang_2016}. This is because of the supper-additivity of coherent information, i.e., for two quantum channels, denoted as $\Phi_1$ and $\Phi_2$, the coherent information of the joint channel $\Phi_1\otimes\Phi_2$ satisfies a specific inequality\cite{shor_quantum_1996,leditzky_platypus_2022,divincenzo_quantum-channel_1998},

  \begin{equation}
  Q_1(\Phi_1\otimes\Phi_2)\ge Q_1(\Phi_1)+Q_1(\Phi_2),
  \end{equation}
this inequality can be  strict\cite{hastings_superadditivity_2009,fern_lower_2008,divincenzo_quantum-channel_1998,leditzky_quantum_2018,bausch_error_2021,siddhu_positivity_2021,smith_additive_2007}. Consequently, bounding techniques are commonly used to obtain upper and lower bounds on the quantum capacity \cite{kianvash1, kianvash2, poshtvan_capacities_2022,fern_lower_2008}. In this paper, we review some bounding techniques that are efficiently computable and then apply them to our channel (\ref{lambdax}). These bounds are not monotone as a function of the parameter $x$. Consequently, we invoke two different bounds from the combination of which we will find a rather narrow region for the quantum capacity of this channel. 
\subsubsection{Semi-definite programming upper bound of quantum capacity}
\label{SDP-Bound-Section}
% Semi-definite Explanation
A general upper bound for quantum capacity using semi-definite programming is introduced in ref.\cite{wang_2016} and is denoted by $Q_\Gamma$. Let  $\Phi:A\rightarrow B$, be a quantum channel and let $ J(\Phi) = \sum_{ij} (\dyad{i}{j}) \otimes (\Phi(\dyad{i}{j}))$ be its Choi matrix \cite{choi1975}. Let $\rho_A$ be an arbitrary density matrix in $A$ and $R$ be an arbitrary positive semi-definite linear operator in $L^+(A\otimes B)$. Then it is shown that  \cite{wang_2016}, 

\be Q(\Phi)\leq Q_\Gamma(\Phi):= \log\left(\max \:\Tr(J(\Phi)R)\right),\ee
where maximization is done on the set of all density matrices $\rho_A$ and all positive semi-definite matrices $R$ subject to the condition 
$$ -\rho_A\otimes I_B \le R^{T_B}\le \rho_A \otimes I_B,$$
where  \(T_B\) represents the partial transpose operation with respect to space \(B\), defined as \((\dyad{ij}{kl})^{T_B} = \dyad{il}{kj}\).
In this paper, we will use this method to upper bound the quantum capacity of the channel (\ref{lambdax}).

\subsubsection{Flagged Extension Upper bound}
Flagged extension is a technique that has proven to be effective for finding upper bounds on the quantum capacity of various quantum channels. It involves constructing a new channel with a higher dimensional output Hilbert space for any channel that can be expressed as a convex combination of other channels. While this technique does not provide a general upper bound and requires specific settings to be tight and computable, it has been successful in investigating the quantum capacities of several channels \cite{smith_additive_2007}\cite{kianvash1}\cite{kianvash2}\cite{kianvash_bounding_2022}. In this section, we use two different flagged extensions and obtain two different upper bounds for the quantum channel which complement each other. \\

\noindent {\bf a: Flagged extension with orthogonal flags}\\

\noindent An upper bound for the quantum capacity of a quantum channel can be found using the following theorem, which is the result of a flagged extension of degradable channels with orthogonal flags. The theorem was first proved in reference \cite{smith_additive_2007}:
\begin{theorem}
\cite{smith_additive_2007} 
Suppose we have
$$\Phi=\sum_i p_i \Phi_i,$$
where $\Phi_i$ is degradable for all $i$. The following inequality holds:
\be
Q(\Phi)\le \sum_i p_iQ(\Phi_i)=\sum_i p_i Q_1(\Phi_i),
\ee
\label{Flagged-Exstension}
where the last equality is due to the degradibility assumption of all  $\Phi_i$s. 
\end{theorem}
\noindent 
The channel $\Lambda_x$ is already in a convex form $\Lambda_x=(1-x)\Lambda_0(\rho)+x\Lambda_1(\rho)$, where $\Lambda_0$ and $\Lambda_1$ are both degradable (see section \ref{DegAndAntiDeg}). This allows a simple upper bound to be found by using the above theorem. Since the $\Lambda_{1}$ channel is known to be anti-degradable (see the discussion after equation (\ref{LS-DAD})), its quantum capacity is zero. We simply obtain

\begin{equation}
\label{flagged_extension_ineq}
Q(\Lambda_x) \leq (1-x) Q_1(\Lambda_0)+x Q_1(\Lambda_1),
\end{equation}

\noindent since $\Lambda_1$ is anti-degradable $Q_1(\Lambda_1)=0$, and $Q_1(\Lambda_0)=0$ being the quantum capacity of the identity channel, is simply found from 
\begin{equation}
\label{identity_map_capacity}
Q_1(\Lambda_0) = \max_{\rho}\: \left[S(\Lambda_0(\rho))-S({\Lambda}^c_0(\rho))\right] = \max_{\rho} \left[S(\rho)-S(\tr(\rho))\right] 
= \max_{\rho} S(\rho) = \log 3,
\end{equation}
which leads to 

\begin{equation}
Q(\Lambda_x) \leq (1-x) Q_1(\Lambda_0) = (1-x) \log 3.
\end{equation}
In Figure \ref{fig:Q-capacity}, we denote this upper bound as $Q_{f1}$.\\

\noindent {\bf b: Flagged extension with non-orthogonal flags}\\

\noindent An alternative degradable flagged extension of the channel (\ref{lambdax}) for $0\le x\le \frac{1}{2}$, can be achieved through \cite{kianvash_bounding_2022}: 

\be
\label{flagged-extensionChannel}
\mathbb{\Lambda}_x(\rho)=(1-x)\rho \otimes \dyad{\phi_0}+x\Lambda_1(\rho)\otimes \dyad{\phi_1},
\ee
where  
\[
\ket{\phi_0}=\sqrt{\frac{1-2x}{1-x}}\ket{0}+\sqrt{\frac{x}{1-x}}\ket{1}, \:\:\:\:\ket{\phi_1}=\ket{0}.
\]
This results in:
\[
Q({\Lambda}_x)\le Q(\mathbb{\Lambda}_x)=Q_1(\mathbb{\Lambda}_x),
\]
with the last equality arising from the degradability of the flagged extension channel. Notably, the covariance of this channel remains $SO(3)$. Let's denote the covariance group of this channel as $G$, and assume that $U_g$ for $g\in G$ is an irreducible representation of this group. By exploiting the irreducible covariance of the channel (\ref{flagged-extensionChannel}), we have:
\be
\label{irreduciblyEQ}
\frac{1}{|G|}\sum_{g\in G} U_g \rho U_g^\dagger=\frac{I}{d},
\ee
where $d$ is the dimension of the input space of the channel (in this case, $d=3$). Moreover, due to the degradability of the flagged extension channel, the coherent information is concave. By leveraging the invariance of the coherent information under the covariance of the channel, we obtain:
\[
I_c(\mathbb{\Lambda}_x,\frac{I}{3})=I_c(\mathbb{\Lambda}_x,\frac{1}{|G|}\sum_{g\in G} U_g \rho U_g^\dagger)\ge \frac{1}{|G|}\sum_{g\in G} I_c(\mathbb{\Lambda}_x, U_g \rho U_g^\dagger)=I_c(\mathbb{\Lambda}_x,\rho),
\]
where $|G|$ is the cardinality of the group $G$. The inequality comes from the concavity of the coherent information, and the equality is due to equation (\ref{irreduciblyEQ}). Hence:

\[
Q_1(\mathbb{\Lambda}_x)=\max_\rho I_c(\mathbb{\Lambda}_x,\rho)=I_c(\mathbb{\Lambda}_x,\frac{I}{3}).
\]
Therefore, it is sufficient to evaluate $I_c(\mathbb{\Lambda}_x,\frac{I}{3})=S(\mathbb{\Lambda}_x(\frac{I}{3}))-S(\mathbb{\Lambda}_x\otimes I\: \dyad{\phi^{+}})$, where $\ket{\phi^{+}}=\frac{1}{3}\sum_{i=0}^3 \ket{ii}$. In this regard, first, we calculate $S(\mathbb{\Lambda}_x(\frac{I}{3}))$:

\[
S(\mathbb{\Lambda}_x(\frac{I}{3}))=S\left(\frac{I}{3} \otimes \left[(1-x)\dyad{\phi_0}+x\dyad{\phi_1}\right]\right)=\log(3)+S\left[(1-x)\dyad{\phi_0}+x\dyad{\phi_1}\right].
\]
Now, we should calculate $S(\mathbb{\Lambda}_x\otimes I \: \dyad{\phi^{+}})$. The nonzero eigenvalues of $J(\mathbb{\Lambda}_x)$ are $(1-x,\frac{x}{3},\frac{x}{3}\frac{x}{3})$ (see appendix \ref{eigenvaluesofflag}), hence:

\[
S(J(\mathbb{\Lambda}_x))=-(1-x)\log(1-x)-x\log(\frac{x}{3}).
\]
We have visualized and labeled this upper bound as $Q_{f2}$ in Figure \ref{fig:Q-capacity}. We have found (not reported here for simplicity) that for our channel and all values of the parameter $x$, the bounds introduced here are lower than the other upper bounds introduced in the literature, like the partial transposition bound \cite{holevo_tanspose_2001}, which is efficiently computable by SDP \cite{watrous_2009,watrous_2012}. Therefore, to obtain a reasonable upper bound for all values of the parameter $x$, we only need to compare the flag-extension upper bounds and the SDP bound. 
\subsection{Lower bound for the quantum capacity}

In order to restrict the allowable values of the quantum capacity, we also need a lower bound. Due to the super-additivity of the channel 
(i.e. $Q_1(\Phi)\leq \frac{1}{n}Q_1(\Phi^{\otimes n})$), a natural lower bound for $Q(\Lambda_x)$ is given by $Q_1(\Lambda_x)$. This is given by 
\begin{equation}
\label{coherent_info_lower_bound}
Q_1(\Lambda_x)=\max_\rho I_c(\Lambda_x,\rho).
\end{equation}
The optimization problem in inequality (\ref{coherent_info_lower_bound}) can be evaluated using various numerical methods. 
%%New explanation
Here, we adopt a hybrid strategy that combines the capabilities of the `fmincon` local optimization algorithm and the GlobalSearch global optimization algorithm, both accessible through MATLAB's Optimization Toolbox and Global Optimization Toolbox. This approach is tailored to effectively address the challenges posed by our non-convex optimization problem. It's worth noting that our problem is mathematically equivalent to minimizing the negative of the coherent information, expressed as $-\min_\rho -I_c(\Lambda_x,\rho)$. Therefore, we address this optimization problem, which necessitates finding minima. GlobalSearch expands the search scope by initially sampling points within the solution space. It then uses `fmincon` to fine-tune and identify local minima in these initial points. \\

\noindent  We can also follow a semi-analytical method which confirms this numerical result. Let us take an ansatz for $\rho$ of the form
$\rho_0=diag (s,1-2s,s)$, where $s$ is a real parameter.  In view of (\ref{coherent_info_lower_bound}), it is true that 

\be\label{ansatz} 
\max_s I_c(\Lambda_x, \rho_0)\leq Q_1(\Lambda_x)\leq Q(\Lambda_x).
\ee
The optimization over this single parameter can then be performed numerically, the result of which coincides with the one performed by the previously mentioned toolbox. We obtain that for $x$ less than $0.38$, $s=\frac{1}{3}$ which means that $\rho_0=\frac{I}{3}$ and for $x$ greater than $0.38$, $I_c(\Lambda, \frac{I}{3})$ becomes negative and hence will not be a meaningful lower bound. For this region, we find the lower bound is actually equal to zero, which is achieved by any of the following pure states $|1\ra:=(1,0,0)^T$, $|0\ra:=(0,1,0)^T, $ and $|-1\ra:=(0,0,1)^T$. In fact one sees from (\ref{lambdax}) and (\ref{Comp:explicit}) that 
\be
\Lambda_x(|1\ra\la 1|)=\begin{pmatrix}1-x&&\\&\frac{x}{2}&\\ &&\frac{x}{2}\end{pmatrix},
\ee
and
\be
\Lambda^c_x(|1\ra\la 1|)=\begin{pmatrix}1-x&&&\\&0&&\\&&\frac{x}{2}&\\ &&&\frac{x}{2}\end{pmatrix},
\ee
both of which have the same entropy thus leading to
$I_c(\Lambda_x,|1\ra\la 1|)=0$. The same thing happens for the other two pure states. This lower bound precisely coincides with the result obtained through numerical search. \\

\begin{figure}[H]
  \centering
    \includegraphics[width=\textwidth]{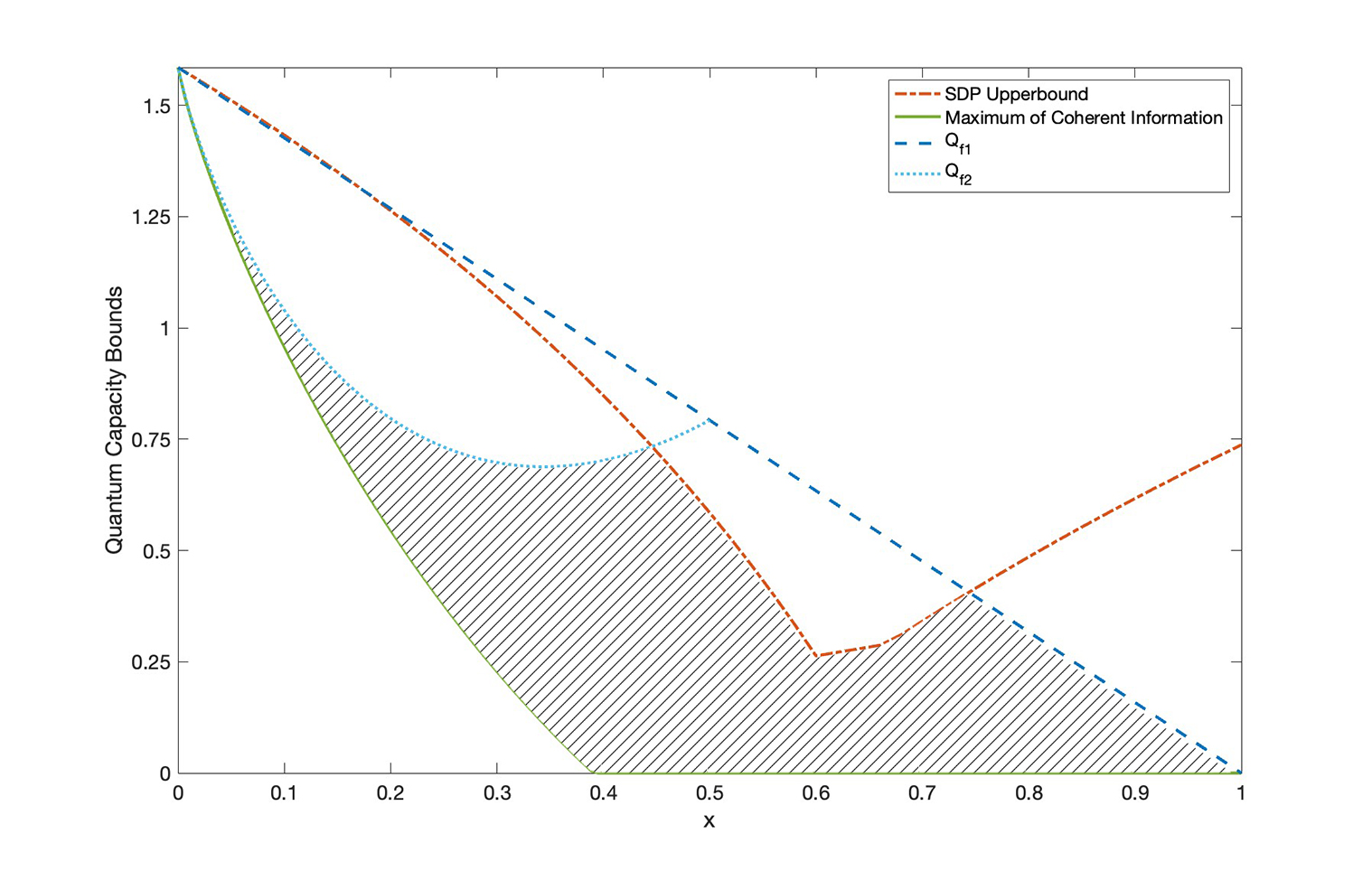}
    \caption{Various bounds for the generalized Landau-Streater Channel are presented. The figure illustrates that the flagged extension upper bound $Q_{f2}$ performs optimally in the region $0\le x \le 0.4435$, while the SDP upper bound excels in the region $0.4435 \le x \le 0.75$. n the remaining regions, the flagged extension upper bound $Q_{f1}$ outperforms other bounds. The hatched area in the figure signifies the potential quantum capacity values. The lower bound for the quantum capacity, the green solid line, is obtained both by a numerical search as in (\ref{coherent_info_lower_bound}) and by taking an ansatz for this optimum state as in (\ref{ansatz}). The results agree. Moreover the lower bound is zero for $x=1$ which conforms with the anti-degradability of the channel $\Lambda_1$.  The parameter $x$ and the quantum capacity are dimensionless.}
    \label{fig:Q-capacity}
  \end{figure}

\noindent Combining the two upper bounds, namely the SDP upper bound and the flag-extension upper bound and the lower bound $Q_1(\Lambda_x$), we find the hashed area in figure (\ref{fig:Q-capacity}) for the allowable values of the quantum capacity of the channel $\Lambda_x$.  

	\section{Anti-degradability of the channel $\Lambda_x$ for $\frac{4}{7}\leq x\leq 1$}\label{ADD}
	It is clear that the Holevo-Werner or the Landau-Streater channel $\Lambda_1$ is both degradable and anti-degradable and hence its quantum capacity is zero. The question arises if these two properties still hold for a certain period near $x=1$. In this section we show that the channel $\Lambda_x$ is anti-degradable in the interval $\frac{4}{7}\leq x\leq 1$. Hence in this interval, the channel $\Lambda_x$ has zero quantum capacity. In view of the superadditivity of the quantum channel and the extreme difficulty of its calculation, this is a significant result. To prove this statement, we find 
 a channel $\mathcal{N}$ such that:
	
	\[
	\Lambda_x(\rho)=\mathcal{N}(\Lambda_x^c(\rho)).
	\]
	
\ni	First let us remind the reader about the fundamental representation of SO(3) generators  that we used and the Kraus operators of the complementary channel $\Lambda_c$. With $k,l$ and $m \in\ \{1,2,3\} $ in cyclic order, we have from (\ref{KrausLSS}) and (\ref{Rs}) that
		\be
	J_m=-i(|k\ra\la l|-|l\ra \la k)\h R_k=\sqrt{1-x}|0\ra\la k|-\sqrt{\frac{x}{2}}J_k	,\ \ \ k=1,2,3.
	\ee
	Note that the Kraus operators $R_k$ are $4\times 3$ dimensional matrices with rows indexed by $0,1,2,$ and $3$ and columns indexed by $1,2$ and $3$. 
	We now 
	define the channel $\mathcal{N}$ by the following $3\times 4$ dimensional Kraus Operators:
	\be
N_0=t\sum_{i=1}^3\dyad{i}=tI_3,\h 	N_k=\frac{1}{\sqrt{3}}|k\ra\la 0|-r J_k,\h k=1,2,3.
\ee
where $I_3$ and $J_k$ are the embedding of the $3-$ dimensional matrices in the corresponding blocks of the matrices, i.e. 
$$N_0=\begin{pmatrix} {\ 0}& tI_3\end{pmatrix}\h  N_k=\begin{pmatrix} \frac{1}{\sqrt{3}}|k\ra&-rJ_k\end{pmatrix}.$$
The parameter $r$ is  chosen to satisfy
\be\label{rx} r\sqrt{\frac{x}{2}}=\sqrt{\frac{1-x}{3}},\ee and $t$ is a real parameter to be determined (note that the phase of a complext $t$ can be always removed to make it real, without affecting the definition of the channel.)
A simple calculation shows that 	
\be
N_0^\dagger N_0+	\sum_{k=0}^3N_k^\dagger N_k=|0\ra\la 0|+(t^2+2r^2)I_3,
\ee
where the requirement of trace-preserving demands that 
\be
t^2+2r^2=1.
\ee
We also need the following simply proved identity  
\be
J_kJ_l=\delta_{kl}I_3-|l\ra\la k|.
\ee
It is then a simple matter to use (\ref{rx}) and show that 
\ba
N_0R_k&=&-t\sqrt{\frac{x}{2}}J_k,\cr
N_kR_k&=&\sqrt{\frac{1-x}{3}}|k\ra\la k|+r\sqrt{\frac{x}{2}}J_k^2=\sqrt{\frac{1-x}{3}}|k\ra\la k|+r\sqrt{\frac{x}{2}}\Big(I_3-|k\ra\la k|\Big)=\sqrt{\frac{1-x}{3}}I_3\cr
N_kR_l&=&\sqrt{\frac{1-x}{3}}|k\ra\la l|+r\sqrt{\frac{x}{2}}J_kJ_l=\sqrt{\frac{1-x}{3}}|k\ra\la l|-r\sqrt{\frac{x}{2}}|l\ra\la k|=i\sqrt{\frac{1-x}{3}}J_{m},
\ea
where we have noted the cyclic order of the indices $(k,\ l,\ m ).$
We can now evaluate $\mathcal{N}(\Lambda_x^c(\rho))$:
\ba \mathcal{N}(\Lambda_x^c(\rho))&=&\sum_{k=0,l=1}^3 N_kR_l(\rho)(N_kR_l)^\dagger\cr
&=&\sum_k (N_0 R_k)(\rho)(N_0R_k)^\dagger+\sum_k (N_k R_k)(\rho)(N_k R_k)^\dagger+\sum_{k\neq l}(N_kR_l)(\rho)(N_kR_l)^\dagger\cr
&=&\frac{x}{2}t^2\sum_k J_k\rho J_k^\dagger+ (1-x)\rho +\frac{2}{3}(1-x)\sum_m J_m\rho J_m^\dagger\cr
&=&(1-x)\rho+\left(\frac{x}{2}t^2+\frac{2}{3}(1-x)\right)\sum_k J_k\rho J_k^\dagger.\ea	
The right hand side will be the same as $\Lambda_x(\rho)$, if $t$ satisfies
\be
\frac{x}{2}t^2+\frac{2}{3}(1-x)=\frac{x}{2}
\ee
which in view of (\ref{rx}) is the same as the trace-preserving condition. This requires that the parameter $t$ satisifes	
\be
t^2=\frac{7x-4}{3x}
\ee
which means that in the region $\frac{4}{7}\leq x\leq 1$, the channel is anti-degradable and hence its quantum capacity is zero.

	\section{Conclusion and Outlook}
Up until now, the Werner-Holevo or the Landau-Streater channel, being an extreme point in the space of qutrit channels has been of interest only due to its structural properties.  We have shown that when suitably modified, this channel can in fact be looked at as a familiar noise model on three-level states. The noise consists of random rotations in different directions by arbitrary angles. In view of the wider interest \cite{karimi1, karimi2, karimi3, karimi4, wang_qudits_2020,hugh_trapped-ion_2005,klimov_qutrit_2003, molina-terriza_experimental_2005,bogdanov_qutrit_2004, groblacher_experimental_2006, lanyon_manipulating_2008,schaeff_experimental_2015,babazadeh_high-dimensional_2017,lu_quantum_2020,bianchetti_control_2010,danilin_experimental_2018,yurtalan_implementation_2020,kononenko_characterization_2021,klimov_qutrit_2003,randall_efficient_2015,baekkegaard_realization_2019,lindon_complete_2023,low_practical_2020,ringbauer_universal_2022} in qutrits as a potential candidate for quantum information processing, our result may find applications in modeling realistic noise on qutrits.  The interesting point is that the action of these continuous random rotations can be represented by three simple Kraus operators. This puts our channel on the same footing for qutrits as the depolarizing channel for qubits.

 The difference with the familiar depolarising channel  is that the latter does not have a simple Kraus representation in terms of physically 
We have then proceeded to determine many of the physical properties of this channel, including its various capacities, where we have found exact expressions for the Holevo quantity, the entanglement-assisted capacity, and upper and lower bounds for its quantum capacity. \\

Our work can be extended in a number of ways. The first one reveals itself, when we look at equations (\ref{Phi}). It is readily seen that the coefficient $a$ cannot be zero, and in fact, the smallest value of this coefficient is $\frac{1}{15}$, which is achieved when all the rotations are equal to $\pm \frac{\pi}{2}$. This is the closest distance that we can come to the Werner-Holevo channel by random unitary rotations. 
The question of how close we can come to an extreme point by random unitary operations is an open question, not only for the Werner-Holevo channel but for any extreme point of CPT maps in general. 
 This problem can be pursued further, if we study the higher spin representations of the Landau-Streater channel,   or even by generalizing the Landau-Streater channel so that we replace the generators of $SO(3)$  with those of $SO(d)$ group. This new channel is completely different from the Landau-Streater channel, i.e., it has $d(d-1)/2$ Kraus operators in $d-$ dimensions but is still equivalent to the Werner-Holevo channel, and consequently it is $SU(d)$ covariant. These are open problems that remain for future publications.   \\
			
\noindent {\Large{\bf Acknowledement:}} This research was supported in part by Iran National Science Foundation, under Grant No.4022322. The authors wish to thank V. Jannessary, M. Nobakht, A. Farmanian, A. Najafzadeh,  A. Tangestani, L. Memarzadeh, A. Rezakhani, S. Oskuie, and M. Mirkamali, for their valuable comments and suggestions.

	\section{Appendix: Protocols for entanglement-assisted classical communication}
	\label{ProtocolsforEAC}
	In this appendix, we propose two different protocols for dense coding via the Werner-Holevo channel (\ref{WHLS}) and calculate the corresponding mutual information. In the first protocol, we let the entangled state $|\Phi\ra_{AB}=\frac{1}{\sqrt{3}}(|00\ra+|11\ra+|22\ra)$ be shared between the two players whom we call Alice (A) and Bob (B). Alice encodes her trit $n=\ 0,\ 1,\ 2$ into the entangled state by acting on her share of the state by $Z^n$, where  $Z=\begin{pmatrix}1&&\\&\omega&\\ &&\omega^2\end{pmatrix}$, where $\omega^3=1$, and in this way encodes her trit $m$ into 
	$$|\Phi_n\ra_{AB}=\frac{1}{\sqrt{3}}(|00\ra+\omega^n|11\ra+\omega^{2n}|22\ra.$$ When her share passes through the channel, Bob will find the complete state to be 
	\ba
	\rho_n&=&(\Lambda_{1}\otimes I)|\Phi_n\ra\la\Phi_n |=\frac{1}{3}(\Lambda_{1}\otimes I)\sum_{j,k}\omega^{(j-k)n}|j,j\ra\la k,k|\cr
	&=&\frac{1}{3}\sum_{j,k}\omega^{(j-k)n}\Lambda_{1}\big(|j\ra\la k|)\otimes |j\ra\la k|\cr
&=&	\frac{1}{6}\sum_{j,k}\omega^{(j-k)n}\big(\delta_{j,k}I-|k\ra\la j|)\otimes |j\ra\la k|\cr
&=&\frac{1}{6}\Big(I\otimes I-\sum_{j,k}\omega^{(j-k)n} |k,j\ra\la j,k|\Big).
	\ea

\noindent Now that Bob has the full state $\rho_n$, he has the task of determining the index $n$ by an optimum measurement. It is not so easy to find this optimal measurement, or maybe there are better protocols for encoding classical trites into entangled states and sending it to Bob. 	In the present protocol, one possible measurement is the following POVM,
	\be
	E_p=\sum_{l=0}^{2}|l,l+p\ra\la l,l+p|\h p=0,\ 1,\ 2.
	\ee
	A straightforward calculation then shows that 
	\be
	P(y=p|x=n)\equiv \tr(E_p\rho_n)=\frac{1}{2}(1-\delta_{p,n}).
	\ee
	This leads to mutual information
	\be
	I(X:Y)=H(X)+H(Y)-H(X,Y)=\log 3+\log 3-\log 6=\log 3-1,
	\ee
	which is below the analytical value $C_{ea,x=1}(\Lambda_x)=\log 3$. Another dense coding protocol is that Alice performs a unitary operator $Z^nX^m$ on the entangled state $|\Phi\ra_{AB}$ and turns it into one of the nine maximally entangled Bell states, hence encoding two classical trits $(m,n)$ into this state and send her own qutrit to Bob (through the channel $\Lambda_{LS}$) who will make a Bell measurement and determine the classical pair of trits $(m,n)$. However, detailed calculation shows that the mutual information in this scenario is again equal to $\log_2 3-1$. Therefore, it is an interesting problem to see what kind of protocol saturates the value $C_{ea}=\log 3.$\\

	\section{Appendix: Eigenvalues of ($\mathbb{\Lambda_x}\otimes I \: \dyad{\phi^+}$)}
	\label{eigenvaluesofflag}
	In this appendix, we evaluate the eigenvalues of the Choi matrix of the flagged extension channel with non-orthogonal flags. To achieve this we expand $(\mathbb{\Lambda_x}\otimes I) \: \dyad{\phi^+}$ as follows:
	$$(\mathbb{\Lambda_x}\otimes I) \: \dyad{\phi^+}=\frac{1}{3}\sum_{i,j=1}^3\left[(1-x)\dyad{i}{j}\otimes \dyad{\phi_0}\otimes\dyad{i}{j}+\frac{x}{2}\sum_{\alpha=1}^3 J_\alpha \dyad{i}{j}J_\alpha \otimes \dyad{\phi_1}\otimes \dyad{i}{j}\right].$$
Since $\dyad{i}{j}$, $\dyad{\phi_0}\otimes\dyad{i}{j}$, $J_\alpha \dyad{i}{j}J_\alpha$ and $\dyad{\phi_1}\otimes \dyad{i}{j}$ are all square matrices there exist a permutation matrix $Q$, such that
$$(\mathbb{\Lambda_x}\otimes I) \: \dyad{\phi^+}=\frac{1}{3}\sum_{i,j=1}^3 Q\left[(1-x) \dyad{\phi_0}\otimes\dyad{i}{j}\otimes \dyad{i}{j}+\frac{x}{2}\sum_{\alpha=1}^3  \dyad{\phi_1}\otimes \dyad{i}{j}\otimes J_\alpha \dyad{i}{j}J_\alpha \right]Q^T $$
Now we define $\ket{J_\alpha}=\frac{1}{2}\sum_i\ket{i}\otimes J_\alpha \ket{i}$, with this definition we can write the above equation in a simpler form.
$$(\mathbb{\Lambda_x}\otimes I) \: \dyad{\phi^+}= Q\left[(1-x) \dyad{\phi_0}\otimes \dyad{\phi^+}+\frac{x}{3}\sum_{\alpha=1}^3 \dyad{\phi_1}\otimes \dyad{J_\alpha} \right]Q^T.$$
Now observe that $\bra{J_\alpha}\ket{J_\beta}=\Tr(J_\alpha^\dagger J_\beta)=\delta_{\alpha\beta}$ and $\bra{J_\alpha}\ket{\phi^+}=\Tr(J_\alpha^\dagger)=0$, which indicates that the eigenvalues of $(\mathbb{\Lambda_x}\otimes I) \: \dyad{\phi^+}$ are $(1-x,\frac{x}{3},\frac{x}{3},\frac{x}{3})$.

	\newpage
	\bibliography{refs}

\end{document}